\documentclass[conference]{IEEEtran}

\usepackage{amsmath,epsfig,bm,amssymb,amsthm}
\usepackage{psfrag}
\usepackage{cite}

\ifCLASSINFOpdf
\else
\fi

\usepackage{amsmath,epsfig,bm,amssymb,amsthm,balance}

\newcommand{\dd}{{\mathrm d}}

\newcommand{\bb}{{\mathrm b}}

\newcommand{\A}{\mathbf{A}}
\newcommand{\Lb}{\mathbf{L}}
\newcommand{\B}{\mathbf{B}}

\newcommand{\hb}{\mathbf{h}}
\newcommand{\Rb}{\mathbf{R}}
\newcommand{\Ib}{\mathbf{I}}
\newcommand{\y}{\mathbf{y}}
\newcommand{\w}{\mathbf{w}}
\newcommand{\diag}{\mbox{diag}}
\newcommand{\dgs}{\mbox{diag}\{\mathbf{s}\}}
\newcommand{\dgsc}{\mbox{diag}\{\mathbf{s}^*\}}
\newcommand{\s}{\mathbf{s}}
\newcommand{\U}{\mathbf{U}}

\newcommand{\Sgw}{\mathbf{\Sigma}_{\w}}
\newcommand{\E}{\mathrm{E}}
\newcommand{\C}{\mathbf{C}}
\newcommand{\CN}{\mathbf{\mathcal{C}}^N}

\newcommand{\dgh}{\mbox{diag}\{\mathbf{h}_2\}}
\newcommand{\dghc}{\mbox{diag}\{\mathbf{h}_2^*\}}

\begin{document}
%
\title{Differential Dual-Hop Relaying over Time-Varying Rayleigh-Fading Channels
\thanks{The authors are with the Department of Electrical and Computer Engineering,
University of Saskatchewan, Saskatoon, Canada, S7N5A9.
Email: m.avendi@usask.ca, ha.nguyen@usask.ca.}}

\author{\IEEEauthorblockN{M. R. Avendi and Ha H. Nguyen}
Department of Electrical and Computer Engineering\\
University of Saskatchewan, Saskatoon, Canada, S7N5A9\\
Email: m.avendi@usask.ca, ha.nguyen@usask.ca.
}

\maketitle

\begin{abstract}
\label{abs}
This paper studies dual-hop amplify-and-forward relaying over time-varying Rayleigh fading channels with differential $M$-PSK modulation and non-coherent detection. For the case of ``two-symbol'' detection, a first order time-series model is utilized to characterize the time-varying nature of the cascaded channel. Based on this model, an exact bit error rate (BER) expression is derived and confirmed with simulation results. The obtained expression shows that the BER is related to the auto-correlation of the cascaded channel and an irreducible error floor exists at high transmit power. To overcome the error floor experienced with fast-fading, a nearly optimal ``multiple-symbol'' differential sphere detection (MSDSD) is also developed. The error performance of MSDSD is illustrated with simulation results under different fading scenarios.
\end{abstract}


\IEEEpeerreviewmaketitle

\section{Introduction}
\label{se:intro}
Dual-hop transmission has been studied in the literature as a technique to leverage coverage problems of many wireless applications such as 3GPP LTE, WiMAX, WLAN, Vehicle-to-Vehicle communication and wireless sensor networks \cite{coop-deploy,coop-V2V,coop-WiMAX,coop-dohler,coop-LTE}. Such a technique can be seen as a type of cooperative communication in which one node in the network helps another node to communicate with (for example) the base station when the direct link is very poor or the user is out of coverage. A commonly used two-phases transmission process is usually utilized. Here the source transmits data to the relay in the first phase, while in the second phase the relay performs a decode-and-forward (DF) or amplify-and-forward (AF) strategy to send the received data to the destination \cite{coop-laneman}. The simplicity of the AF strategy makes it attractive for many applications. Error performance of dual-hop relaying using AF strategy over \emph{slow-fading} channels has been studied in \cite{dual-hop-Hasna}. Also, the statistical properties of the cascaded channel between the source and destination in a dual-hop AF network have been examined in \cite{SPAF-P}.

This paper studies dual-hop transmission over \emph{time-varying} Rayleigh fading channels which uses differential $M$-PSK modulation at the transmitter, AF strategy with a fixed gain at the relay, and non-coherent detection at the destination. We refer to this system as differential dual-hop (D-DH). Differential modulation with non-coherent detection is employed to avoid any channel estimation at the relay or destination. The non-coherent detection can be carried out over either two-symbol duration or multiple-symbol duration. For the case of two-symbol detection, it is well known that over slow-fading channels, around 3 dB loss is seen between coherent and non-coherent detections. However, in practical time-varying channels, the effect of channel variation can lead to a much larger degradation. To evaluate this loss, based on the first-order autoregressive model, AR(1), of the individual Rayleigh-faded channels \cite{AR1-ch}, a time-series model, proposed originally in \cite{DAF-ITVT,DAF-WCNC}, is utilized to characterize the time-varying nature of the equivalent channel. With the two-symbol non-coherent detection, an exact bit error rate (BER) expression is obtained. It is also shown that an error floor exists (at high transmit power), which is related to the auto-correlation of the cascaded channel. 

Since the two-symbol non-coherent detection does not perform well over fast time-varying channels, a near optimal multiple-symbol differential sphere detection (MSDSD) scheme is developed. The MSDSD was first proposed for point-to-point systems in \cite{MSDSD-L} to reduce the complexity of multiple-symbol differential (MSD) detection. In the context of relay networks, due to the complexity of the distribution of the received signal at the destination, the optimal decision rule of MSD decoding cannot be easily obtained for the D-DH system under consideration. Instead, an alternative decision rule is proposed and its near optimal performance is illustrated with simulation results. 
It should also be mentioned that, based on the Gaussian assumption for the received signal, the authors in \cite{MSDSD-Hanzo} developed a MSDSD scheme for multi-node differential AF networks.

The outline of the paper is as follows. Section \ref{sec:system} describes the system model. In Section \ref{sec:two-symbol}, two-symbol differential detection and its performance over time-varying channels is considered. Section \ref{sec:MSDSD} develops the MSDSD algorithm. Simulation results are given in Section \ref{sec:sim}. Section \ref{sec:con} concludes the paper.

\emph{Notation}: Bold upper-case and lower-case letters denote matrices and vectors, respectively. $(\cdot)^t$, $(\cdot)^*$, $(\cdot)^H$ denote transpose, complex conjugate and Hermitian transpose of a complex vector or matrix, respectively. $|\cdot|$  denotes the absolute value of a complex number and $\|\cdot \|$ denotes the Euclidean norm of a vector. $\mathcal{CN}(0,N_0)$ stands for complex Gaussian distribution with zero mean and variance $N_0$. $\mbox{E}\{\cdot\}$ denotes expectation operation. Both ${e}^{(\cdot)}$ and $\exp(\cdot)$ show the exponential function. $\dgs$ is the diagonal matrix with components of $\s$ on the
main diagonal and $\Ib_N$ is the $N \times N$ identity matrix. A symmetric $N\times N$ Toeplitz matrix is defined by $\mbox{toeplitz}\{x_1,\cdots,x_N\}$. $\mbox{det}\{\cdot\}$ denotes determinant of a matrix. $\CN$ is the set of complex vectors with length $N$.

\section{System Model}

\label{sec:system}
The wireless relay model under consideration has one source, one relay and one destination. The direct link between the source and the destination is assumed to be very weak and hence not used. Therefore, the source has to communicate with the destination via the relay. Each node has a single antenna, and the communication between nodes is half duplex (i.e., each node is able to only send or receive in any given time). The channels from the source to the relay and from the relay to the destination are denoted by $h_1[k]\sim \mathcal{CN}(0,1)$ and $h_2[k]\sim \mathcal{CN}(0,1)$, respectively, where $k$ is the symbol time. A Rayleigh flat-fading model is assumed for each channel. The channels are spatially uncorrelated and changing continuously in time. The time-correlation between two channel coefficients, $n$ symbols apart, follows the Jakes' model: 
\begin{equation}
\varphi_i(n)=\E \{h_i[k]h_i^*[k+n]\}=J_0(2\pi f_i n),\quad i=1,2
\end{equation}
where $J_0(\cdot)$ is the zeroth-order Bessel function of the first kind and $f_i$ is the maximum normalized Doppler frequency of the $i$th channel.

At time $k$, a group of $\log_2M$ information bits is mapped to a $M$-PSK symbol as $v[k]\in \mathcal{V}$ where $\mathcal{V}=\{e^{j2\pi m/M},\; m=0,\dots, M-1\}$. Before transmission, the symbols are encoded differentially as
$
\label{eq:s-source}
s[k]=v[k] s[k-1],\quad s[0]=1.
$
The transmission process is divided into two phases. Block-by-block transmission protocol is utilized to transmit a frame of symbols in each phase as symbol-by-symbol transmission
causes frequent switching between reception and transmission, which is not practical. However, the analysis is the same for both cases and only the channel auto-correlation value is different ($n=1$ for block-by-block and $n=2$ for symbol-by-symbol). In phase I, the symbol $\sqrt{P_0}s[k]$ is transmitted by the source to the relay, where $P_0$ is the average source power per symbol. The received signal at the relay is
\begin{equation}
\label{eq:relay_rx}
x[k]=\sqrt{P_0}h_1[k]s[k]+w_1[k]
\end{equation}
where $w_1[k]\sim \mathcal{CN}(0,N_0)$ is the noise at the relay.

The received signal at the relay is then multiplied by an amplification factor $A$, and re-transmitted to the destination. The common amplification factor, based on the variance of SR channel, is commonly used in the literature as $A =\sqrt{P_1/(P_0+N_0)}$, where $P_1$ is the average power per symbol at the relay. However,
$A$ can be any arbitrarily fixed value. The corresponding received signal at the destination is
\begin{equation}
\label{eq:dest-rx1}
y[k]=Ah_2[k]x[k]+w_2[k],
\end{equation}
where $w_2[k]\sim \mathcal{CN}(0,N_0)$ is the noise at the destination. Substituting (\ref{eq:relay_rx}) into (\ref{eq:dest-rx1}) yields
\begin{equation}
\label{eq:destination-rx}
y[k]= A \sqrt{P_0}h[k]s[k]+w[k],
\end{equation}
where $h[k]=h_1[k]h_2[k]$ is the cascaded channel, and
$
w[k]=A h_2[k]w_1[k]+w_2[k]
$
is the equivalent noise at the destination.



\section{Two-Symbol Differential Detection}
\label{sec:two-symbol}

\subsection{Time-Series Model and Differential Detection}
\label{subsec:channel-model}
The conventional two-symbol differential detection was developed under the assumption that $h[k]\approx h[k-1]$. However, such an assumption is not valid for fast time-varying channels. To find the performance of two-symbol differential detection in time-varying channels, individual channels are expressed by an AR(1) model as
\begin{gather}
\label{eq:ARi}
h_i[k]=\alpha_i h_i[k-1]+\sqrt{1-\alpha_i^2} e_i[k],\quad i=1, 2
\end{gather}
where $\alpha_i=\varphi_i(1) \leq 1$ is the auto-correlation of the $i$th channel and $e_i[k]\sim \mathcal{CN}(0,1)$ is independent of $h_i[k-1]$. Based on these expressions, a first-order time-series model has been derived in \cite{DAF-WCNC} to characterise the evolution of the cascaded channel in time. The time-series model of the cascaded channel is given as (the reader is referred to \cite{DAF-WCNC} for the detailed derivations/verification)
\begin{equation}
\label{eq:ARmodel}
h[k]=\alpha h[k-1]+\sqrt{1-\alpha^2}\ h_2[k-1]e_2[k]
\end{equation}
where $\alpha=\varphi_1(1)\varphi_2(1) \leq 1$ is the equivalent auto-correlation of the cascaded channel, which is equal to the product of the auto-correlations of individual channels \cite{SPAF-P}, and $e_2[k]\sim \mathcal{CN}(0,1)$ is independent of $h[k-1]$. 

By substituting (\ref{eq:ARmodel}) into (\ref{eq:destination-rx}) one has
\begin{equation}
\label{eq:cddfast}
y[k]=\alpha v[k]y[k-1]+n[k],
\end{equation}
where
\begin{multline}
\label{eq:n[k]}
n[k]=w[k]- \alpha v[k]w[k-1]
\\+ \sqrt{1-\alpha^2}A\sqrt{P_0} s[k] h_2[k-1] e_2[k].
\end{multline}

Finally, the two-symbol differential detection is expressed as
\begin{equation}
\label{eq:ml-detection}
\hat{v}[k]= \arg \min \limits_{v[k]\in \mathcal{V}} |y[k]- v[k] y[k-1]|^2
\end{equation}

\subsection{BER Performance Analysis}
\label{sec:symbol_error_probability}
As can be seen from the model in \eqref{eq:n[k]}, the exact distribution of $n[k]$ is difficult to find. However, by substituting $h_2[k]$ from \eqref{eq:ARi} into $w[k]$ one has
\begin{equation}
\label{eq:w[k]-hat}
w[k]
\approx w_2[k]+A \alpha_2 w_1[k]  h_2[k-1]
\end{equation}
where the approximation comes from the observation that even for very fast-fading channels the term $A\sqrt{1-\alpha_2^2}$ is very small. Hence, using the approximated value of $w[k]$ into \eqref{eq:n[k]},
\begin{multline}
\label{eq:n[k]-app}
n[k]\approx w_2[k]+A\alpha_2 w_1[k] h_2[k-1]\\
-\alpha v[k] \left( w_2[k-1]+Ah_2[k-1]w_1[k-1] \right)\\
+\sqrt{1-\alpha^2} A \sqrt{P_0} s[k] h_2[k-1] e_2[k]
\end{multline}
which shows that, conditioned on $h_2[k-1]$, $n[k]$ is a combination of complex Gaussian random variables and hence it is also complex Gaussian random variable. From now on, the time index $[k-1]$ is omitted in this section to simplify the notation.

Using the unified approach in \cite[eq.25]{unified-app}, it follows that the conditional BER for differential modulation can be written as
\begin{equation}
\label{eq:Pb-gama-hrd}
P_{\bb}(E|\gamma,h_2)=\frac{1}{4\pi} \int \limits_{-\pi}^{\pi} g(\theta) e^{-q(\theta)\gamma} \dd \theta
\end{equation}
where $g(\theta)=(1-\beta^2)/(1+2\beta\sin(\theta)+\beta^2)$, $q(\theta)=(b^2/\log_2 M) (1+2\beta\sin(\theta)+\beta^2)$, and $\beta=a/b$. The values of $a$ and $b$ depend on the modulation size \cite{unified-app}. Also, $\gamma$ is the instantaneous effective SNR at the output of the differential detector.

For time-varying channels, based on \eqref{eq:cddfast} and \eqref{eq:n[k]-app}, by dividing the signal power to the noise power, $\gamma$ is defined as
\begin{equation}
\label{eq:gama_d}
\gamma=\bar{\gamma} |h_1|^2
\end{equation}
where $$\bar{\gamma}=\frac{\alpha^2 A^2 (P_0/N_0) |h_2|^2}{(\alpha_2^2+\alpha^2)(1+A^2|h_2|^2)+(1-\alpha^2)A^2 (P_0/N_0) |h_2|^2}.$$
It can be seen that for slow-fading $(\alpha=1,\;\alpha_2^2=1)$, $\gamma$ is half of that of the coherent value $A^2 (P_0/N_0) |h_2|^2 |h_1|^2/(1+A^2|h_2|^2) $ \cite{dual-hop-Hasna,SPAF-P}. This causes the so-called 3 dB performance loss between coherent and non-coherent detections. For time-varying channels $(\alpha < 1)$, the effect of the term $(1-\alpha^2)A^2 (P_0/N_0) |h_2|^2$ in the denominator of $\bar{\gamma}$ leads to a larger degradation in the effective SNR and overall  performance.

Since, $|h_1|^2$ is exponentially distributed, $\gamma$, conditioned on $|h_2|$, follows an exponential distribution  with pdf of
\begin{equation}
\label{eq:pdf-gama_b}
f_{\gamma|h_2}=\frac{1}{\bar{\gamma}} \exp\left(-\frac{\gamma}{\bar{\gamma}}\right)
\end{equation}

By substituting $\gamma$ into \eqref{eq:Pb-gama-hrd} and taking the average over the distribution of $\gamma$, one has
\begin{equation}
\label{eq:Pb-hrd}
P_{\bb}(E|h_2)=\frac{1}{4\pi} \int \limits_{-\pi}^{\pi} g(\theta) I(\theta) \dd \theta
\end{equation}
where
\begin{multline}
\label{eq:I_theta}
I(\theta)=\int \limits_{0}^{\infty} e^{-q(\theta)\gamma} \frac{1}{\bar{\gamma}} e^{-\frac{\gamma}{\bar{\gamma}}} \dd \gamma=\frac{1}{\bar{\gamma}q(\theta)+1}\\
=b_3(\theta) \frac{\lambda+b_1}{\lambda+b_2(\theta)}
\end{multline}
with $\lambda=|h_2|^2$, $b_3(\theta)=b_2(\theta)/b_1$ and $b_1$, $b_2(\theta)$ defined as
\begin{multline}
\label{eq:b1b2b3}
b_1=\frac{1+\alpha^2}{(\alpha_2^2+\alpha^2)A^2+(1-\alpha^2)A^2(P_0/N_0)}\\
b_2(\theta)=\frac{1+\alpha^2}{(1+\alpha^2)/b_1+q(\theta)A^2(P_0/N_0)}
\end{multline}

Now, by taking the final average over the distribution of $|h_2|^2$, $f_{\lambda}=e^{-\lambda}$, it follows that
\begin{equation}
\label{eq:Pb}
P_{\bb}(E)=\frac{1}{4\pi} \int \limits_{-\pi}^{\pi} g(\theta) J(\theta) \dd\theta
\end{equation}
where
\begin{multline}
\label{eq:J_theta}
J(\theta)=\int \limits_{0}^{\infty} b_3(\theta) \frac{\lambda+b_1}{\lambda+b_2(\theta)} e^{-\lambda} \dd\lambda\\
=b_3(\theta) \left( 1+(b_1-b_2(\theta)) e^{b_2(\theta)} E_1(b_2(\theta)) \right)
\end{multline}
and $E_1(x)=\int \limits_x^{\infty} (e^{-t}/t) \dd t$ is the exponential integral function. The definite integral in \eqref{eq:Pb} can be easily computed using numerical methods and it gives the exact value of the BER of the D-DH system under consideration in time-varying Rayleigh fading channels.

It is also informative to examine the expression of $P_{\bb}(E)$ at high transmit power. In this case, $\lim \limits_{(P_0/N_0)\rightarrow \infty} E[\bar{\gamma}]= \alpha^2/(1-\alpha^2)$ which is independent of $|h_2|^2$ and $(P_0/N_0)$. Therefore, by substituting the converged value into (\ref{eq:Pb-hrd}), the error floor appears as
\begin{equation}
\label{eq:PEP-floor}
\lim \limits_{(P_0/N_0) \rightarrow \infty} P_{\bb}(E)= \frac{1}{4\pi} \int \limits_{-\pi}^{\pi} g(\theta) \frac{1-\alpha^2}{\alpha^2q(\theta)+1-\alpha^2} \dd \theta
\end{equation}
This error floor will be observed in the simulation results.

\section{Multiple-Symbol Detection}
\label{sec:MSDSD}
As discussed in the previous section, two-symbol non-coherent detection suffers from a high error floor in fast time-varying channels. To overcome such a limitation, this section develops a multiple-symbol detection scheme that takes a window of the received symbols at the destination for detecting the transmitted signals.

Let the $N$ received symbols be collected in vector $\y=\left[\; y[1],y[2],\dots, y[N]\; \right]^t$, which can be written as
\begin{equation}
\label{eq:Y}
\y=A\sqrt{P_0} \dgs \diag\{\hb_2\} \hb_1 +\w
\end{equation}
where $\s= \left[\; s[1],\cdots, s[N]\; \right]^t$, $\hb_2=\left[\; h_2[1],\cdots, h_2[N] \;\right]^t$, $\hb_1=\left[\; h_1[1],\cdots, h_1[N]\; \right]^t$ and $\w=\left[\; w[1],\cdots, w[N]\; \right]^t$.
Therefore, conditioned on both $\s$ and $\hb_2$, $\y$ is a circularly symmetric complex Gaussian vector with the following pdf:
\begin{equation}
\label{eq:pdfY}
P(\y|\s,\hb_2)=\frac{1}{\pi^N \mathrm{det}\{\Rb_{\y}\}} \exp\left( -\y^H \Rb_{\y}^{-1} \y \right).
\end{equation}
In \eqref{eq:pdfY}, the matrix $\Rb_{\y}$ is the conditional covariance matrix of $\y$, defined as
\begin{equation}
\label{eq:RY}
\begin{split}
\Rb_{\y}=&\E \{ \y \y^H | \s,\hb_2 \}=\\
&A^2 P_0 \dgs \diag\{\hb_2\} \Rb_{\hb_1} \diag\{\hb_2^*\} \dgsc+\Sgw
\end{split}
\end{equation}
with $$\Rb_{\hb_1}=\E\{ \hb_1 \hb_1^H \}=\mathrm{toeplitz}\{ 1,\varphi_1(1),\dots,\varphi_1(N-1)\},$$  $$\Sgw=N_0 \mbox{diag}\{\left(1+A^2 |h_2[1]|^2\right),\cdots,\left(1+A^2|h_2[N]|^2\right) \}$$ as the covariance matrices of $\hb_1$ and $\w$, respectively.

Based on \eqref{eq:pdfY}, the maximum likelihood (ML) detection would be given as
\begin{equation}
\label{eq:ML}
\hat{\s}=\arg \max \limits_{\s \in \CN} \left\lbrace \underset{\hb_2}{\E} \left\lbrace
\frac{1}{\pi^N \mathrm{det}\{\Rb_{\y}\}} \exp\left( -\y^H \Rb_{\y}^{-1} \y \right)
\right\rbrace \right\rbrace.
\end{equation}
where $\hat{\s}= \left[\; \hat{s}[1],\cdots, \hat{s}[N] \; \right]^t$.
As it can be seen, the ML metric needs the expectation over the distribution of $\hb_2$, which does not yield a closed-form expression. 
As an alternative, it is proposed to use the following modified decision metric:
\begin{equation}
\label{eq:ML-Modified}
\hat{\s}=\arg \max \limits_{\s \in \CN} \left\lbrace
\frac{1}{\pi^N \mathrm{det}\{\bar{\Rb}_{\y}\}} \exp\left( -\y^H \bar{\Rb}_{\y}^{-1} \y \right)
\right\rbrace
\end{equation}
where
\begin{multline*}
\bar{\Rb}_{\y}=\underset{\hb_2}{\E} \{ \Rb_{\y} \}=A^2P_0 \dgs \Rb_{\hb} \dgsc\\+(1+A^2)N_0 \Ib_N
=\dgs\; \C \; \dgsc
\end{multline*}
\begin{equation}
\label{eq:C}
\C=A^2P_0  \Rb_{\hb} +(1+A^2)N_0\Ib_N
\end{equation}
\begin{multline}
\label{eq:R_h}
\Rb_{\hb}=\E\left\lbrace \dgh \Rb_{\hb_1} \dghc \right\rbrace=\\
\mathrm{toeplitz} \{ 1,\varphi_1(1)\varphi_2(1),\dots,\varphi_1(N-1)\varphi_2(N-1) \}.
\end{multline}
Although, the simplified decision metric is not optimal in the ML sense, it will be shown by simulation results that nearly identical performance to that obtained with the optimal metric can be achieved. 

Using the rule $\det \{\A\B\}=\det\{\B\A\}$, the determinant in \eqref{eq:ML-Modified} is no longer dependent to $\s$ and the modified decision metric can be further simplified as
\begin{multline}
\label{eq:ML-simp}
\hat{\s}=\arg \min \limits_{\s\in \CN} \left\lbrace \y^H \bar{\Rb}_{\y}^{-1} \y \right\rbrace\\
=\arg \min \limits_{\s\in \CN} \{\y^H \dgs \C^{-1} \dgsc \y \}\\
=\arg \min \limits_{\s \in \CN} \left\lbrace (\diag\{\y\} \s^*)^H \Lb\Lb^H \diag\{\y\} \s^* \right\rbrace\\
=\arg \min \limits_{\s\in \CN} \left\lbrace  \|\U \s \|^2 \right\rbrace
\end{multline}
where $\U=(\Lb^H \diag\{\y\})^*$ and $\Lb$ is obtained by the Cholesky decomposition of
$\C^{-1}=\Lb \Lb^H$.


The minimization in \eqref{eq:ML-simp} can then be solved using the MSDSD function described in \cite{MSDSD-L} with low complexity. The MSDSD algorithm adapted to the D-DH system under consideration is summarized in \emph{Algorithm I} as MSDSD-DH. It should be mentioned that steps 1 to 3 are calculated once, whereas steps 4 to 6 will be repeated for every $N$ consecutive received symbols.

\begin{table}[thb!]
\renewcommand{\arraystretch}{2}
\label{tb:msdsd}
\centering
\begin{tabular}{l}
\hline
\bf Algorithm 1: MSDSD-DH \\
\hline
{\bf Input:} $f_1, f_2, A, P_0,N_0, M, N, \y$ \\
{\bf Output:} $\hat{v}[k],\quad k=1,\cdots,N-1$ \\
\hline
1: Find $\Rb_{\hb}$ from \eqref{eq:R_h} \\
2: Find $\C$ from \eqref{eq:C} \\
3: Find $\Lb$ from $\C^{-1}=\Lb\Lb^H$  \\
4: Find $\U=(\Lb^H \diag\{\y\})^*$ \\
5: Call function $\hat{\s}$=MSDSD ($\U$,$M$) \cite{MSDSD-L}\\
6: $\hat{v}[k]=\hat{s}^*[k] \hat{s}[k+1],\quad k=1,\cdots,N-1$\\
\hline
\end{tabular}
\end{table}


\section{Simulation Results}
\label{sec:sim}
In this section the dual-hop relay network under consideration is simulated in different scenarios. 
In all simulations, the channel coefficients, $h_1[k]$ and $h_2[k]$, are generated based on the simulation method of \cite{ch-sim}. This simulation method was developed to generate channel coefficients that are correlated in time. The amount of time-correlation is determined by the normalized Doppler frequency of the underlying channel, which is a function of the speed of the vehicle, carrier frequency and symbol duration. Obviously, for fixed carrier frequency and symbol duration, a higher vehicle speed leads to a larger Doppler frequency and less time-correlation.

Based on the normalized Doppler frequencies of the two channels, different cases can be considered. In Case I, it is assumed that all nodes are fixed or slowly moving so that both channels are slow-fading with the normalized Doppler values of $f_1=.001$ and $f_2=.001$. In Case II, it is assumed that the source is fast moving so that the SR channel is fast-fading with $f_1=.01$. On the other hand, the relay and destination are fixed and the RD channel is slow-fading with $f_2=.001$. In Case III, it is assumed that both the source and the destination are fast moving so that both the SR and RD channels are fast-fading with $f_1=.02$ and $f_2=.01$, respectively.

\begin{figure}[htb!]
\psfrag {P(dB)} [][b] [1]{$(P_0/N_0)$ (dB)}
\psfrag {BER} [] [] [1] {BER}
\psfrag {Case I} [] [b] [.8] {Case I}
\psfrag {Case II} [] [] [.8] {Case II}
\psfrag {Case III} [] [t] [.8] {Case III}
\psfrag {Error Floors} [] [t] [.8] {Error Floors}
\centerline{\epsfig{figure={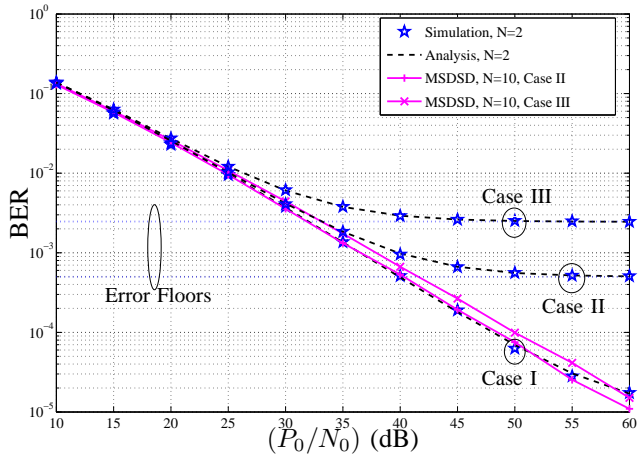},width=8.5cm}}
\caption{BER of a D-DH network in different fading cases using DBPSK and non-coherent detection with $N=2$ and $N=10$ symbols.}
\label{fig:ber_m2}
\end{figure}
\begin{figure}[htb!]
\psfrag {P(dB)} [][b] [1]{$(P_0/N_0)$ (dB)}
\psfrag {BER} [] [] [1] {BER}
\psfrag {Case I} [] [] [.8] {Case I}
\psfrag {Case II} [] [] [.8] {Case II}
\psfrag {Case III} [] [] [.8] {Case III}
\psfrag {Error Floors} [] [t] [.8] {Error Floors}
\centerline{\epsfig{figure={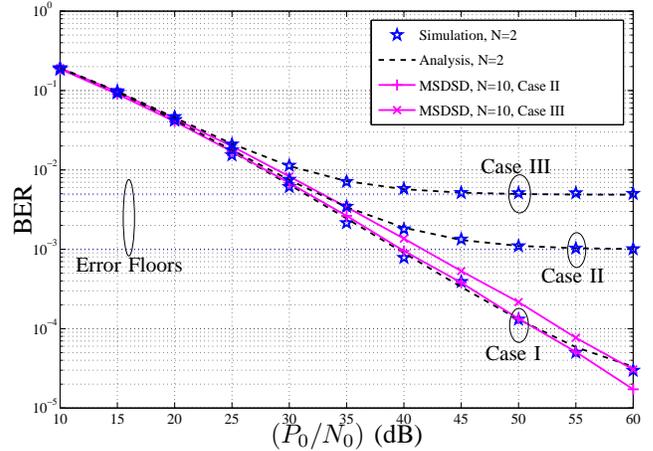},width=8.5cm}}
\caption{BER of a D-DH network in different fading cases using DQPSK and non-coherent detection with $N=2$ and $N=10$ symbols.}
\label{fig:ber_m4}
\end{figure}

In each case, information bits are differentially encoded with either BPSK ($M=2$) or QPSK ($M=4$) constellations. Although an arbitrary power allocation between the source and the relay can be used, equal power allocation, namely $P_0=P_1$, is assumed where $P_1$ is the relay power. The amplification factor at the relay is fixed to $A=\sqrt{P_1/(P_0+N_0)}$ to normalize the average relay power to $P_1$. At the destination, first, the two-symbol differential non-coherent detection is applied. The simulation is run for various values of the source power. The simulated BER values are computed for all cases and are plotted versus the source power in Figs.~\ref{fig:ber_m2} and ~\ref{fig:ber_m4}, for DBPSK and DQPSK, respectively.

For evaluating the theoretical BER values when the two-symbol detection is applied, the value of $\alpha$ is computed for each case. Also, $\{a=0, b=\sqrt{2}\}$, $\{a=\sqrt{2-\sqrt{2}}, b=\sqrt{2+\sqrt{2}}\}$, are obtained for DBPSK and DQPSK, respectively \cite{unified-app}. The corresponding theoretical BER values and the error floors are computed from (\ref{eq:Pb}) and \eqref{eq:PEP-floor} and plotted in the figures with dashed and dotted lines, respectively.

As can be seen in Fig.~\ref{fig:ber_m2}, in case I, the BER is monotonically decreasing with $(P_0/N_0)$ and it is consistent with the theoretical values in (\ref{eq:Pb}). However, the plot starts to flat out after $(P_0/N_0)=55$ dB, which means that it reaches an error floor at very high SNR (which is practically insignificant). In Case II, which involves one fast-fading channel, this phenomena starts earlier, around $35$ dB, and leads to an error floor at $5\times 10^{-4}$, which can also be predicted from (\ref{eq:PEP-floor}). The performance degradation is much more severe after $25$ dB in Case III since both channels are fast-fading, which leads to an error floor at $3\times 10^{-3}$. Similar behaviours can be seen in Fig. ~\ref{fig:ber_m4} when using DQPSK modulation. As is clearly seen in both Figs.~\ref{fig:ber_m2} and ~\ref{fig:ber_m4}, the simulation results verify our theoretical evaluations.

Given the poor performance of the two-symbol detection in Cases II and III, MSDSD-DH algorithm is applied which takes a window of $N=10$ symbols for detection. With known Doppler values of the channels, ${\Rb}_{\hb}$ and then $\Lb$ are computed. Then for $N$ consecutive received symbols, the upper triangular matrix $\U$ is found and given to the MSDSD function described in \cite{MSDSD-L} to recover ($N-1$) transmitted symbols. The BER results of the MSDSD-DH algorithm are also plotted in Figures~\ref{fig:ber_m2} and ~\ref{fig:ber_m4} with solid lines (different legends). Since, the best performance is achieved in the slow-fading environment, the performance plot of Case I can be used as a benchmark to see the effectiveness of MSDSD-DH. It can be seen that the MSDSD-DH is able to bring the performance of the system in Case II and Case III very close to that of Case I.

\section{Conclusion}
\label{sec:con}
Differential AF relaying for a dual-hop transmission has been studied for time-varying fading channels. The obtained BER expression for the two-symbol detection shows that the error performance depends on the overall fading rate of the equivalent channel and an error floor exists at high SNR. For fast-fading channels a large fading rate leads to a severe performance degradation of the two-symbol detection. A near optimal multi-symbol detection algorithm was also presented to improve the performance of the system in fast time-varying channels.


\bibliographystyle{IEEEbib}
\bibliography{h:/latex/references}

\end{document}